\magnification=\magstep1
\tolerance=500
\rightline{30  July, 2017}
\bigskip
\centerline{\bf Relativistic Entanglement}
\bigskip
\centerline{Lawrence Horwitz$^{1,2,3}$\footnote{*}{larry@post.tau.ac.il}}
\centerline{and}
\centerline{Rafael I. Arshansky$^4$}
\bigskip
 \centerline{${}^1$ Tel Aviv University, Ramat Aviv, 69978 Israel}
\centerline{${}^2$ Ariel University, Ariel, 40700 Israel}
\centerline{${}^3$ Bar Ilan University, Ramat Gan, 52900 Israel}
\centerline{${}^4$ Etzel Street 12/14 HaGiva HaZorfatit, Jerusalem, 9785412 Israel}
\bigskip
\noindent{Abstract}
\par  The relativistic quantum theory of Stueckelberg, Horwitz and Piron (SHP) describes in a simple way the experiment on interference in time of an electron emitted by femtosecond laser pulses carried out by Lindner {\it et al}. In this paper, we show that, in a way similar to our study of the Lindner {\it et al} experiment (with some additional discussion of the covarant quantum mechanical description of spin and angular momentum), the experiment proposed by Palacios {\it et al} to demonstrate entanglement of a two electron state, where the electrons are separated in time of emission, has a consistent interpretation in terms of the SHP theory. We find, after a simple calculation, results in essential agreement with those of Palacios {\it et al}; but with the observed times as values of  proper quantum observables.
\bigskip
\bigskip
\noindent{\bf 1. Introduction}
\par Palacios, Rescigno and McCurdy [1] have described a proposed experiment
which could show entanglement of a two electron system in which each electron is emitted at a slightly different time. Although the anticipation of this effect is very reasonable, it does not have a theoretical justification in the framework of the standard nonrelativistic quantum theory, since in the nonrelativistic theory, both electrons must be prepared in states at precisely eual times. As for the Lindner {\it et al} [2] experiment showing interference in time for the wave function of a particle, for which extensive calculations were done using the nonrelativistic Schr\"odinger evolution of the electron, wave functions at different times (corresponding to elements of different Hilbert spaces [3]) are incoherent in the nonrelativistic quantum theory. The direct product states corresponding to the basis for many body systems must, in the same way, be constructed from states in the same Hilbert space. Therefore, the  same conclusion can be reached for the entanglement of the spins of a two body system. In actual practice, in fact, it would not be possible experimentally to generate two body states at precisely equal times, so that it is important to construct a theoretical basis, as we shall do below, in which effects of the type we expect to see (and are seen, for example, in the experiment of Lindner {\it et al} [2]) can be consistently described.
\par The nonrelativistic theory of the two body state with spin is constructed from linear combinations of 
direct product wave functions taken at equal time [4]. One could argue intuitively from the vector model, in which the result ${\bf J}^2 = j(j+1)$ (for ${\bf J}$ the angular momentum operator, and $j$ the integer or half-integer eigenvalue), that it appears that the physical angular momentum is not precisely along the ``direction'' of the vector ${\bf J}$, but can be thought of as precessing around it. The entangled spin zero state of two spin $1/2$ systems therefore would be the result of an exact synchronization of these oppositely oriented precessing spins so that the total angular momentum is zero. At slightly different times, this synchronization would be, in principle, lost. Under nonrelativistic Schr\"odinger evolution the superposition of two-body states at different times would therefore be ineffective.  Stated more rigorously, states are not coherent [3] at nonequal times and linear superposition is not defined in the nonrelativistic theory.
\par As for the Lindner {\it et al} experiment[2], an explanation can be given in terms of the relativistic quantum theory of Stueckelberg, Horwitz and Piron (to be called SHP) [5]. The computation in terms of the SHP [6] was in precise agreement with the experiment result (actually predicted in 1976 [7], when the technology was not available for verification). In this paper, we apply a similar reasoning to the entangled two body state.
\par We start with a review of the basic SHP theory[5] and a discussion of how the Wigner theory of induced representations for relativistic spin is applied in this framework. We then argue that the proposed experiment of Palacios {\it et al} should yield well-defined entanglement for the constitutent particle at not precisely equal times.
\par Stueckelberg [5], in 1941, imagined that a particle world line would be straight for no interaction, but that interaction could bend the world line so that it would turn to propagate i n the negative direction of time. To describe such a picture, he introduced an invariant parameter along the world line, which he called $\tau$, and interpreted the backward in time evolving branch of the line as an antiparticle. Horwitz and Piron [5] then generalized this idea in the sense that the parameter $\tau$ was to be considered as a universal invariant time, as for the original postulate of Newton, in order to formulate the many body problem in this framework, as we discuss below.       
\par As a model for the structure of the dynamical laws that might be
considered, Stueckelberg proposed a Lorentz invariant 
Hamiltonian for free motion of the form
$$ K = {p^\mu p_\mu \over 2M}, \eqno(1.1)   $$
 where $M$ is considered a parameter, with dimension mass, associated
 with the particle being described, but is not necessarily its measured
 mass. In fact, the numerator (with metric $-+++$; we generally take $c=1$), 
$$ p^\mu p_\mu = -m^2, \eqno(1.2)$$
corresponds to the actual observed mass (according to the Einstein
relation $ E^2 = {\bf p}^2 + m^2$),
where, in this context, $m^2$ is a dynamical variable. 
\par The Hamilton equations, generalized covariantly to four dimensions, 
are then
$$\eqalign{{\dot x}^\mu\equiv
{dx^\mu \over d\tau} &= {\partial K \over \partial p_\mu}\cr
{\dot p}_\mu\equiv {d p_\mu \over d\tau}&=- {\partial K \over \partial
x^\mu}.
 \cr}\eqno(1.3)$$
\par These equations are postulated to hold for any Hamiltonian model,
 such as with additive potentials or gauge fields. A Poisson bracket 
may be then defined in the same way as for the nonrelativistic theory.  The
construction is as follows. Consider the $\tau$ derivative of a
function $F(x,p)$, {\it i.e.},
$$ \eqalign{{dF\over d\tau} &= {\partial F \over \partial x^\mu}{dx^\mu \over
d\tau} + {\partial F \over \partial p^\mu} {dp^\mu \over d\tau}\cr
&=  {\partial F \over \partial x^\mu} {\partial K \over \partial
p_\mu} - {\partial F \over \partial p^\mu}{\partial K \over \partial
x_\mu}\cr
&= \{F,K\},\cr} \eqno(1.4)$$ 
thus defining a Poisson bracket $\{ F,G\}$ quite generally.  The
arguments of the nonrelativistic theory then apply,{\it i.e.}, that functions
which obey the Poisson algebra isomorphic to their group algebras will
have vanishing Poisson bracket with the Hamiltonian which has the
symmetry of 
that group,and 
are thus conserved quantities, and the (time independant) Hamiltonian itself is then
(identically) a
conserved quantity.
\par It follows from the Hamilton equations that for the free  particle
case 
 $$ {\dot x}^\mu = {p^\mu \over M} \eqno(1.5)$$
and therefore, dividing the space components by the time components,
cancelling the $d\tau$'s ($p^0 = E$ and $x^0=t$),
$$ {d{\bf x} \over dt } = {{\bf p} \over E}, \eqno(1.6)$$
the Einstein relation for the observed velocity. Furthermore, we see
that
$$ {\dot x}^\mu {\dot x}_\mu = {p^\mu p_\mu \over M^2}; \eqno(1.7)$$
with the definition of the invariant
$$ds^2 = -dx^\mu dx_\mu, \eqno(1.8)$$
corresponding to proper time squared (for a timelike interval),
this becomes
$$ {ds^2 \over d\tau^2} = {m^2 \over M^2}. \eqno(1.9)$$
Therefore, the proper time interval $\Delta s$ of a particle along a trajectory
parametrized by $\tau$ is equal to the corresponding interval $\Delta
\tau$ only if $m^2 = M^2$, a condition we shall call ``on mass
shell''.
\par Stueckelberg [5] formulated the quantized version of this theory by postulating the
commutation relations 
$$ [x^\mu, p^\nu] = i \hbar g^{\mu\nu}, \eqno(1.10)$$
where  $g^{\mu\nu}$ is the Lorenrtz metric given above, 
and a Schr\"odinger type equation (we shall take $\hbar =1$ in the following)
$$i {\partial \over \partial \tau} \psi_\tau(x) = K \psi_\tau(x), \eqno(1.11)$$
where $\psi(x)$ is an element of a Hilbert space on $R^4$ satisfying
$$ \int |\psi(x)|^2 d^4x = 1, \eqno(1.12)$$
and satisfies the required Hilbert space property of linear superposition. With the generalization of Horwitz and Piron [5], Eq. $(1.11)$ can be written for any number $N$ of particles as
$$i {\partial \over \partial \tau} \psi_\tau(x_1, x_2 \dots x_N) ) = K \psi_\tau(x_1, x_2 \dots x_N), \eqno(1.13)$$
where $K$ could have, for example, the form
$$ K= \Sigma_i^N {{p_i}^\mu{ p_i}_\mu \over 2M_i} + V(x_1, x_2 \dots x_N), \eqno(1.14)$$
and $V(x_1, x_2 \dots x_N)$ is assumed, for our present purposes, to be Poincar\'e invariant.
\par The basis of the Hilbert space describing such states is provided by the direct product of one particle wave functions {\it taken at equal} $\tau$ (as for equal time $t$ in the nonrelativistic theory [4]). In the following, we apply this structure to the description of two particles with spin.

\bigskip
\noindent{\it 2. Relativistic spin and the Dirac representation}
 \par We shall discuss in this section  the basic idea of a
 relativistic particle with spin, based on Wigner's seminal
 work [8].  The
 theory is adapted here to be applicable to relativistic quantum
 theory; in this form, Wigner's theory, together with the requirements
 imposed by the observed correlation between spin and statistics in
 nature for identical particle systems, makes it possible to define
 the total spin of a state of a relativistic many body system.  
\par  The spin of a particle in a nonrelativistic framework
corresponds to the 
lowest dimensional nontrivial representation of  the rotation 
group; the generators are the Pauli matrices $\sigma_i$ divided by
two, the generators of the fundamental representation of
the double 
covering of $SO(3)$. The self-adjoint operators that are the
generators of this group measure angular momentum and are
associated with magnetic moments. Such a description is not relativistically
covariant, 
but Wigner [8] has shown how to describe this dynamical property of
a particle in a covariant way. The method developed
by Wigner 
provided the foundation for what is now known as the theory of induced 
representations [9], with very wide applications, including a
very powerful approach to finding the representations of noncompact groups [9].
\par In the nonrelativistic quantum theory, the spin states of a two
or more particle system  are defined by combining the spins of these
particles at equal time using appropriate Clebsch-Gordan coefficients
[4][10] at each value of the time.  The restriction to equal time
follows from the tensor product form of the representation of the
quantum states for a  many body problem [4][11]. 
 For two spin $1/2$
(Fermi-Dirac) particles, for example, an antisymmetric space distribution would
correspond to a symmetric combination of the spin factors, {\it i.e.}
a spin one state, and a symmetric space distribution would correspond
to an antisymmetric spin combination, a spin zero
state.   This
correlation is the source of the
famous Einstein-Podolsky-Rosen discussion [11]. The experiment
proposed by Palacios {\it et al} [1] suggests that spin 
entanglement could
occur for two particles at non-equal times; the spin carried by wave
functions of SHP type would naturally carry such correlations over the
width in $t$ of the wave packets, and therefore would provide a simple and rigorous prediction for this experiment.
\par  Wigner's formulation [8], however,  was not appropriate for
application to a consistent relativistic quantum theory, since it does not preserve, as we shall
explain below, the covariance of the expectation value of coordinate
operators [5]. Before constructing a generalization of Wigner's method
which is useful in relativistic quantum theory we first review Wigner's
method in its original form, and show how the difficulties arise.
\par To establish some notation and the basic method,  we start with
the basic principle of relativistic covariance for a scalar quantum
wave function $\psi(p)$.  In a new Lorentz frame described by the
parameters $\Lambda$ of the Lorentz group, for which $p'^\mu =
\Lambda^\mu_\nu p^\nu$ (we work in momentum space here for
convenience), the same physical point in momentum space described in
different coordinates,  by arguing that the probability density  must
be the same, 
$$ \psi'(p') = \psi(p)  \eqno(2.1)$$
up to a phase, which we take to be unity.
It then follows that as a function of $p$, 
$$ \psi'(p) = \psi(\Lambda^{-1} p). \eqno(2.2)$$
Since, in Dirac's notation, 
$$ \psi'(p) \equiv <p|\psi'>, \eqno(2.3)$$
Eq. $(2.2)$ follows equivalently by writing
$$ |\psi'> = U(\Lambda) |\psi> \eqno(2.4)$$
so that
$$ \eqalign{<p|\psi'> &=  <p|U(\Lambda) |\psi> \cr &= <\Lambda^{-1}
 p|\psi>\cr
 &= \psi(\Lambda^{-1}p), \cr}\eqno(2.5)$$
where we have used 
$$U(\Lambda)^\dagger |p> = U(\Lambda^{-1}) |p> = |\Lambda^{-1}p>.\eqno(2.6)$$
 
\par To discuss the transformation properties of the representation of
a relativistic particle with  spin, Wigner proposed that we consider a special
frame in which $p_0^\mu = (m,0,0,0)$; the subgroup of the Lorentz
group that leaves this vector invariant is clearly $O(3)$, the
rotations in the three space in which ${\bf p} =0$, or its covering
$SU(2)$. Under a Lorentz boost, transforming the system to its
representation in a moving inertial frame, the rest momentum 
 appears as $p_0^\mu  \rightarrow p^\mu$, but under this unitary 
transformation, the subgroup that leaves $p_0^\mu$ invariant is
carried to a form which leaves $p^\mu$ invariant, and the group
remains $SU(2)$.   The $2 \times 2$ matrices representing this group
are altered by the Lorentz transformation, and are functions of the
momentum $p^\mu$.   The resulting state then transforms by a further
change in $p^\mu$ and an $SU(2)$ transformation compensating for this
change.  This additional transformation is called the ``little group''
of Wigner.  The family of values of $p^\mu$ generated by Lorentz 
transformations on $p_0^\mu$ is called the ``orbit'' of the induced 
representation. This $SU(2)$, in its lowest dimensional
representation, parametrized by $p^\mu$ and the  additonal Lorentz 
transformation $\Lambda$, corresponds to Wigner's covariant relativistic 
definition of the spin of a relativistic particle [8].
\par We now apply this method to review Wigner's construction based on a
representation induced on the momentum $p^\mu$. Let us define the 
momentum-spin ket
$$ |p,\sigma> \equiv U(L(p))|p_0 ,\sigma>, \eqno(2.7)$$
where $U(L(p))$ is the unitary operator inducing a Lorentz
transformation of the timelike $p_0=(m,0,0,0)$ (rest frame momentum)
to the general timelike vector $p^\mu$. The effect of a further Lorentz
transformation parameterized by $\Lambda$, induced by
$U(\Lambda^{-1})$, can be written as 
$$ U(\Lambda^{-1})|p,\sigma> =
U(L(\Lambda^{-1}p))U^{-1}(L(\Lambda^{-1}p))
U(\Lambda^{-1}) U(L(p)) |p_0, \sigma> \eqno(2.8)$$
The product of the last three unitary factors
$$U^{-1}(L(\Lambda^{-1}p))U(\Lambda^{-1}) U(L(p))
\eqno(2.9)$$
has the property that under this combined unitary transformation, the
ket is transformed so that $p_0\rightarrow p_0$, and thus corresponds
to just a rotation (called the Wigner rotation), the stability
subgroup of the vector $p_0$.  This
rotation can be represented by a $2\times 2$ matrix  acting on the
index $\sigma$, {\it i.e.}, so that
$$ U(\Lambda^{-1})|p,\sigma> =U(L(\Lambda^{-1}p))|p_0, \sigma'>
D_{\sigma, \sigma'}(\Lambda, p)= |\Lambda^{-1}p, \sigma'>D_{\sigma,
\sigma'}(\Lambda, p). \eqno(2.10)$$
where, as a representation of rotations, $D$ is unitary. Therefore,
taking the complex conjugate of
$$ <\psi| U(\Lambda^{-1})|p,\sigma>= <\psi|\Lambda^{-1}p, \sigma'>D_{\sigma,
\sigma'}(\Lambda, p), $$
one obtains 
$$<p,\sigma| U(\Lambda) \psi> =  <\Lambda^{-1}p, \sigma'|\psi>
D_{\sigma',\sigma } (\Lambda p), \eqno(2.11)$$
where, in this construction,  
$$ D_{\sigma', \sigma}(\Lambda,p) = \bigl( (L(p)^{-1}\Lambda
L(\Lambda^{-1}p) )\bigr)_{\sigma',\sigma}\ \  , \eqno(2.12)$$
expressed in terms of the $SL(2,C)$
matrices corresponding to the unitary transformation $(2.9)$.
The result $(2.11)$ can be written as
$$ \psi'(p,\sigma) = \psi(\Lambda^{-1}p, \sigma')D_{\sigma',
\sigma}(\Lambda,p). \eqno(2.13)$$
\par The algebra of the $2\times 2$ matrices of the fundamental
representation of the group $SL(2,C)$ are isomorphic to that of the
Lorentz group, and the product of the corresponding matrices provide
the $2 \times 2$ matrix representation of $D_{\sigma',
\sigma}(\Lambda,p)$; we may therefore have
$$ D_{\sigma', \sigma}(\Lambda,p) = \bigl(L^{-1}(p)\Lambda
L(\Lambda^{-1}p) \bigr)_{\sigma',\sigma}, \eqno(2.14)$$
 where $L$ and $\Lambda$ are the $2 \times 2$ matrices of
 $SL(2,C)$.
\par As we have mentioned above, the presence of the $p$-dependent
matrices representating the spin of a relativistic particle in the
transformation law of the wave function destroys the covariance, in a
relativistic quantum theory, of the expectation value of the 
coordinate operators.  To see this, consider the expectation value of
the dynamical variable $x^\mu$, {\it i.e.}
$$ <x^\mu> = \int d^4 p \psi(p)^\dagger i{ \partial \over \partial
p_\mu} \psi(p).\eqno(2.15) $$ 
\par A Lorentz transformation would introduce the $p$-dependent
$2\times 2$ 
unitary transformation  on the function $\psi(p)$, and the derivative
with  respect to momentum would destroy the covariance property that
we would wish to see of the expectation value $<x^\mu>$. 

\par It is also not possible, in this framework, to form wave
packets of definite spin
by integrating over the momentum variable, since this would add 
functions over different parts of the orbit, with a different $SU(2)$
at each point.
\par As we describe in the following, these problems can be solved 
 by inducing a representation of the spin on a
timelike unit vector $n^\mu$ in place of the four-momentum, 
 using a representation induced on a timelike vector, say, $n^\mu$,
which is independent of $x^\mu$ or $p^\mu$ [12][13]. This
solution also permits the linear superposition of momentum states to
form wave packets of definite spin, and admits the construction of
definite spin states for many body relativistic systems. In the following, we show how such a
representation can be constructed.
\par  Let us
define, as in $(2.7)$,
$$ |n,\sigma, x> \equiv U(L(n)) |n_0, \sigma, x>,
\eqno(2.16)$$
where we may admit a dependence on $x$ (or, through Fourier transform,
on $p$). Here, we distinguish the action of $U(L(n))$ from the general
Lorentz transformation $U(\Lambda)$;  $U(L(n))$ acts only on the
vector space of the $n^\mu$.  Its infinitesimal generators are given
by
$$ M^{\mu\nu}_n = -i (n^\mu {\partial \over \partial n_\nu} - n^\nu 
{\partial \over \partial n_\mu}),
\eqno(2.17)$$
while the generators of the transformations $U(\Lambda)$ act on the
full vector space of both the $n^\mu$ and the $x^\mu$ (as well as
$p^\mu$).  In terms of the canonical variables,  
$$M^{\mu\nu} = M^{\mu \nu}_n + (x^\mu p^\nu - x^\nu p^\mu). \eqno(2.18)$$
The operator $(2.17)$ is self-adjoint in the full Hilbert space norm defined by the integral of the norm (in the sheets of the foliation defined by $n^\mu$) to be defined in $(2.25)$ over $d^4 n  \delta (n^\mu n_\mu +1) d^4 x = {d^3{\bf n}\over n_0}d^4 x $.
The two terms of the full generator commute.
Following the method outlined above, we now investigate the properties
of a total Lorentz transformation, {\it i.e.}
$$ U(\Lambda^{-1}) |n,\sigma, x> = U(L(\Lambda^{-1}n)
 \bigl(U^{-1}(L(\Lambda^{-1}n))U(\Lambda^{-1})U(L(n)))\bigr)
|n_0, \sigma, x>, \eqno(2.19)$$
Now, consider the conjugate of $(2.19)$, 
$$ <n,\sigma, x|U(\Lambda)= <n_0, \sigma,
x|\bigl(U(L^{-1}(n))U(\Lambda)U(L(\Lambda^{-1}n))\bigr)
U^{-1}(L(\Lambda^{-1}n)). \eqno(2.20) $$
\par  The operator in the first factor (in parentheses) preserves $n_0$, and 
therefore, as before, contains an element of the little group associated
with $n^\mu$ which may be represented by the matrices of $SL(2,C)$. 
It also acts, due to the factor  $U(\Lambda)$( for which the
generators are those of the Lorentz group acting both on $n$ and $x$
(or $p$), as in $(2.18)$), taking $x
\rightarrow \Lambda^{-1} x $ in the conjugate ket on the left. Taking
the product on both sides with $|\psi>$, we obtain
$$ <n,\sigma, x| \psi>' = <\Lambda^{-1}n, \sigma', \Lambda^{-1}x|\psi>
D_{\sigma', \sigma}(\Lambda, n), \eqno(2.21)$$
or 
$$ \psi'_{n,\sigma}(x)= \psi_{\Lambda^{-1}n, \sigma'}(\Lambda^{-1} x)
D_{\sigma', \sigma}(\Lambda, n).
\eqno(2.22)$$
 where 
$$ D(\Lambda, n) = L^{-1}(n)\Lambda L(\Lambda^{-1}n), \eqno(2.23)$$
with $\Lambda$ and $L(n)$ the corresponding
$2\times 2$ matrices of $SL(2,C)$
  ($\Lambda$ and $L(n)$ are the corresponding
$2\times 2$ matrices of $SL(2,C)$).
\par With this transformation law, one may take the
Fourier transform to obtain the wave function in momentum space, and 
conversely.  The matrix $D$ is an element of $SU(2)$, and therefore
linear superpositions over momenta or coordinates
maintain the definition of the particle spin, and interference
phenomena for relativistic particles with spin may be studied
consistently. Furthermore, if two
or more particles with spin are represented in representations induced
on $n^\mu$, at a given value of $n^\mu$ on their respective orbits,
their spins can be added by the standard methods with the use of
Clebsch-Gordan coefficients [10]. This method therefore admits the
treatment of a many body relativistic system with spin [14]. It is  interesting to note that the little group rotations defined by $(2.23)$ are in a spacelike surface defined by $n^\mu$. The vector $n^\mu$ may be thought of as the normal to the spacelike surfaces defined by Schwinger [21] in the  discussion of his variational principle for quantum field theory, thus providing a natural framework for the development of a covariant spinor formalism without reference to the momentum representation.       
\par  There are two
fundamental representations of $SL(2,C)$ which are
inequivalent [15]. Multiplication by the operator
$\sigma\cdot p$ of a two dimensional spinor
representing one of these results in an object transforming like the
second representation. Such an operator could be expected to occur in a
dynamical theory, and therefore the state of lowest dimension in
spinor indices of a physical system should contain both
representations [5]. As we shall emphasize, however,  in our treatment of the more
than one particle system, for the rotation subgroup, both of the
fundamental representations yield the same $SU(2)$ matrices up to a
unitary transformation, and therefore the Clebsch-Gordan decomposition
of the product state into irreducible representations may be carried
out independently of which fundamental $SL(2,C)$ representation is
associated with each of the particles [14].
\par  We now discuss the construction
of Dirac spinors. 
\par The defining relation for the fundamental
$SL(2,C)$ matrices is 
 $$ \Lambda^\dagger \sigma^\mu n_\mu \Lambda = \sigma^\mu
 (\Lambda^{-1} n)_\mu, \eqno(2.24)$$
 where $\sigma^\mu = (\sigma^0, {\bf \sigma})$; $\sigma^0$ is the unit
 $2\times 2$ matrix, and ${\bf \sigma}$  are the Pauli matrices. Since
 the determinant of $\sigma^\mu n_\mu$ is the Lorentz invariant
 ${n^0}^2 - {\bf n}^2$, and the determinant of $\Lambda$ is unity in
 $SL(2,C)$, the transformation represented on the left hand side of
 $(2.24)$ must induce a Lorentz transformation on $n^\mu$.  The inequivalent second fundamental
 representation may be constructed by using this defining relation
 with $\sigma^\mu$ replaced by ${\underline\sigma}^\mu\equiv
 (\sigma^0, -{\bf \sigma})$. For every Lorentz transformation $\Lambda$
 acting on $n^\mu$, this defines an $SL(2,C)$ matrix
 $\underline{\Lambda}$ (we use the same symbol for the Lorentz
 transformation on a four-vector as for the corresponding $SL(2,C)$
 matrix acting on the $2$-spinors). 
\par Since both fundamental representations of $SL(2,C)$ should occur
 in the general quantum wave function representing the state of the
 system, the norm in each $n$-sector of the Hilbert space must be defined as
$$ N =  \int d^4x ( |{\hat \psi}_n(x)|^2 + |{\hat \phi}_n(x)|^2), \eqno(2.25)$$
where ${\hat \psi}_n$ transforms with the first $SL(2,C)$ and ${\hat
\phi}_n$
 with the second. From the construction of the little group $(2.21)$, it
follows that $L(n)\psi_n$ transforms with $\Lambda$, and ${\underline
L}(n) \phi_n$ transforms with ${\underline \Lambda}$; making this
replacement in $(2.23)$, and using the fact, obtained from the
defining relation $(3.22)$, that ${L(n)^\dagger}^{-1}L(n)^{-1}= \mp
\sigma^\mu n_\mu$ and ${{\underline L}(n)^\dagger}^{-1}{\underline
L}(n)^{-1}= \mp{\underline \sigma}^\mu n_\mu$, one finds that
$$ N = \mp \int d^4x {\bar \psi} _n (x) \gamma\cdot n \psi_n(x),
\eqno(2.26)$$
where $\gamma \cdot n \equiv \gamma^\mu n_\mu$ (for which $(\gamma
\cdot n)^2 = -1)$, and the matrices $\gamma^\mu$ are the Dirac matrices
as defined in the books of Bjorken and Drell [16].  Here, the
four-spinor $\psi_n(x)$  is defined by
$$ \psi_n(x) = { 1 \over \sqrt{ 2}} \left(\matrix{1&1\cr -1 &
1\cr}\right) \left(\matrix{L(n) {\hat \psi}_n(x) \cr {\underline L}(n)
{\hat \phi}_n(x) \cr}\right), \eqno(2.27)$$
and the sign $\mp$   corresponds to $n^\mu$ in the positive or
negative light cone. The wave function defined in $(2.26)$ transforms as 
$$ \psi'_n (x) = S(\Lambda) \psi_{\Lambda^{-1}n} (\Lambda^{-1} x)
\eqno(2.28)$$
and $S(\Lambda)$ is a (nonunitary) transformation generated
infinitesimally, as in the standard Dirac theory (see, for example [16], by $\Sigma^{\mu\nu} \equiv { i \over 4}
[\gamma^\mu, \gamma^\nu]$. 
 \par The Dirac operator $\gamma\cdot p$ is not Hermitian in the
 (invariant) scalar product associated with the norm $(2.16)$. It is
 of interest to consider the Hermitian and anti-Hermitian parts                
$$\eqalign{ K_L &= { 1\over 2}(\gamma \cdot p + \gamma \cdot n \gamma
\cdot p \gamma \cdot n) = -(p\cdot n) (\gamma \cdot n) \cr 
K_T &= {1\over 2} \gamma^5 (\gamma \cdot p - \gamma\cdot n \gamma\cdot
p \gamma\cdot n) = -2i \gamma^5 (p\cdot K)(\gamma\cdot n),\cr}
\eqno(2.29)$$
where $K^\mu = \Sigma^{\mu\nu}n_\nu $, and we have introduced the
factor $\gamma^5 = i\gamma^0\gamma^1\gamma^2\gamma^3$, which
anticommutes with each $\gamma^\mu$ and has square $-1$ so that $K_T$
is Hermitian and commutes with the Hermitian $K_L$. Since 
$$ K_L^2 = (p\cdot n)^2 \eqno(2.30)$$
and
$$ K_T^2 = p^2 + (p\cdot n)^2, \eqno(2.31)$$
we may consider 
$$ K_T^2 - K_L^2 = p^2 \eqno(2.32)$$
to  pose an eigenvalue problem
analogous to the second order  mass eigenvalue condition for the free
Dirac equation (the Klein Gordon condition). For the Stueckelberg
equation of evolution
corresponding to the free particle, we may therefore take
$$ K_0 = {1 \over 2M} (K_T^2 - K_L^2)={1 \over 2M}p^2 . \eqno(2.33)$$
In the presence of electromagnetic interaction, gauge invariance under
a spacetime dependent gauge transformation (we discuss the more
general case of a gauge transformation depending on $\tau$ as well in
the next chapter), the expressions for $K_T$ and $K_L$ given in $(2.29)$,
in gauge covariant form, then imply, in place of $(2.33)$, 
$$K = {1 \over 2M}(p-eA)^2 + {e \over 2M} \Sigma_n^{\mu\nu}
F_{\mu\nu}(x), \eqno(2.34)$$
where 
$$ \Sigma_n^{\mu\nu}= \Sigma^{\mu\nu} + K^\mu n^\nu -K^\nu n^\mu
 \equiv
{i \over 4} [\gamma_n^\mu, \gamma_n^\nu], \eqno(2.35)$$
where the $\gamma_n^\mu$ are defined in $(2.39)$.
The expression  $(2.34)$ is quite similar to that of the second order Dirac
operator; it is, however, Hermitian in the scalar product defined by $(2.26)$; it has  no direct {\it electric}
coupling to the electromagnetic field in the special frame for which
$n^\mu = (1,0,0,0)$ in the minimal coupling model we have given here
(note that in his calculation of the anomalous magnetic moment,
Schwinger [18] puts the electric field to zero; a non-zero electric field
would lead to a non-Hermitian term in the standard Dirac propagator,
the inverse of the Klein-Gordon square of the interacting Dirac
equation).  The matrices
$\Sigma_n^{\mu\nu}$ are, in fact, a relativistically covariant form
of the Pauli matrices.
\par To see this, we note that the quantities $K^\mu$ and
$\Sigma_n^{\mu \nu}$ satisfy the commutation relations 
 $$\eqalign{ [K^\mu,K^\nu] &= -i \Sigma_n^{\mu\nu}\cr
[\Sigma_n^{\mu\nu}, K^\lambda] &= -i[(g^{\mu\lambda} + n^\nu n^\lambda)
K^\mu - (g^{\mu\lambda} + n^\mu n^\lambda) K^\nu, \cr
[\Sigma_n^{\mu\nu}, \Sigma_n^{\lambda\sigma}] &= -i[(g^{\nu\lambda} +
n^\nu n^\lambda)\Sigma_n^{\mu\sigma} +(g^{\sigma\mu} + n^\sigma n^\mu)
\Sigma_n^{\lambda\nu} \cr
&-(g^{\mu\lambda} + n^\mu n^\lambda)\Sigma_n^{\nu\sigma} +
(g^{\sigma\nu} + n^\sigma n^\nu) \Sigma_n^{\lambda\nu}].\cr}
\eqno(2.36)$$
Since $K^\mu n_\mu = n_\mu\Sigma_n^{\mu\nu} = 0$, there are only three
independent $K^\mu$ and three $\Sigma_n^{\mu\nu}$. The matrices 
$\Sigma_n^{\mu\nu}$ are a covariant form of the Pauli matrices, and the last of
$(3.34)$ is the Lie algebra of $SU(2)$ in the spacelike surface
orthogonal to $n^\mu$. The three independent $K^\mu$ correspond to
the non-compact part of the algebra which, along with the
$\Sigma_n^{\mu\nu}$ provide a representation of the Lie algebra of the
full Lorentz group.  The covariance of this representation follows from
$$ S^{-1} (\Lambda) \Sigma_{\Lambda n}^{\mu\nu}S(\Lambda)
\Lambda_\mu^\lambda \Lambda_\nu^\sigma = \Sigma_n^{\lambda\sigma} . 
\eqno(2.37)$$
\par In the special frame for which  $n^\mu = (1,0,0,0))$,
$\Sigma_n^{i,j}$ become the Pauli matrices ${1\over 2} \sigma^k$
with $(i,j,k)$ cyclic, and $\Sigma_n^{0j} = 0$. In this frame there is
no direct electric interaction with the spin in the minimal coupling
model $(3.33)$.  We remark that there is, however, a natural spin
coupling which becomes pure electric in the special frame, given by
$$ i[K_T,K_L] = -ie \gamma^5 (K^\mu n^\nu - K^\nu n^\mu) F_{\mu
\nu}. \eqno(2.38)$$
It is easy to see that the value of this commutator
 reduces to $\mp e {\bf \sigma \cdot E}$ in the special frame for
 which $n^0 = -1$; this operator is Hermitian and would correspond to
 an electric dipole interaction with the spin.
 \par Note that the matrices 
$$ \gamma_n^\mu = \gamma_\lambda \pi^{\lambda \mu}, \eqno(3.37)$$
 where the projection
$$ \pi^{\lambda \mu}= g^{\lambda\mu} + n^\lambda n^\mu ,\eqno(3.38)$$
appearing in $(2.36)$,
 play an important role in the description of the dynamics in the induced
representation. In $(2.34)$, the existence of projections on each index
in the spin coupling term implies that $F^{\mu\nu}$ can be replaced by
  ${F_n}^{\mu\nu}$ in this term, a tensor projected into the foliation 
subspace.
\par We further remark that in relativistic scattering theory, the
$S$-matrix is Lorentz invariant (Bjorken (1964)). The asymptotic states can be
decomposed according to the conserved projection operators
$$ \eqalign{P_{\pm}&= {1 \over 2} (1 \mp \gamma\cdot n)\cr
P_{E\pm} &= {1\over 2} (1 \mp {p\cdot n \over |p\cdot n|})\cr
&{\rm and}\cr
P_{n\pm} &= {1 \over 2}(1 \pm {2i\gamma^5 K\cdot p \over [p^2 + (p\cdot
n)^2]^{1/2}}). \cr}\eqno(2.41)$$
The operator
 $$ {2i\gamma^5 K\cdot p \over [p^2 + (p\cdot
n)^2]^{1/2}}\rightarrow {\bf \sigma \cdot p}/|{\bf p}| \eqno(2.42)$$
when $n^\mu \rightarrow (1,0,0,0)$. {\it i.e.}, $P_{n\pm}$ corresponds
to a helicity projection.Therefore the matrix elements of the
$S$-matrix at any point on the orbit of the induced representation is
equivalent (by replacing $S$ by $U(L(n))S U^{-1}(L(n))$) to the
corresponging helicity representation associated
with the frame in which $n^\mu$ is $n^0$

\par The anomalous magnetic
moment of the electron can be computed in this framework (Bennett [19]) 
without appealing to the full quantum field theory of electrodynamics.

\bigskip
\noindent{\it 3 The many body problem with spin, and spin-statistics}
\bigskip
\par As in the nonrelativistic quantum theory, one represents the
state of an $N$-body system in terms of a basis given by the tensor
product of $N$ one-particle states, each an element of a one-particle
Hilbert space. The general state of such an
$N$-body system is given by a linear superposition over this
basis [4]. Second quantization then corresponds
to the construction of a Fock space, for which the set of all $N$
body states, for all $N$,  are imbedded in a large Hilbert space, for
which operators that change the number $N$ are defined [4]. In order to
construct the tensor product space corresponding to the many-body
system, we must consider, as for the nonrelativistic theory, only the product of
wave functions which are elements of the same Hilbert space.  In the
nonrelativistic theory, this corresponds to functions at equal time;
in the relativistic theory, the functions are taken to be at equal
$\tau$. Thus, in the relativistic theory, there are correlations at
unequal $t$, within the support of the Stueckelberg wave
functions. Moreover, for particles with spin we argue that in the
induced representation, these function must be taken at {\it identical
values of $n^\mu$}, i.e., taken at the same point on the orbits of the
induced representation of each particle [20].
\par This statement lies in the observation that the
spin-statistics relation appears to be a universal fact of nature.  An
elementary proof, for example, for
a system of two spin $1/2$ particles, is that a $\pi$ rotation of the
system introduces
a phase factor of $e^{i {\pi \over 2}}$ for each particle, thus
introducing a 
minus sign for the two body state.
However, the $\pi$ rotation is equivalent to an interchange of the two
identical particles.  This argument rests on the fact that each
particle is in {\it the same representation} of $SU(2)$, which can only be
achieved in the induced representation with the particles at the
same point on their respective orbits. The same argument applies for
bosons, which must be symmetric under interchange (in this case the
phase of each factor in a pair is $e^{i\pi}$).
 We therefore see that
identical particles must carry the same value of $n^\mu$, and the
construction of the $N$-body system must follow this rule. It
therefore follows that the two body relativistic system can carry a
spin computed by use of the usual Clebsch-Gordan coefficients, and
entanglement would follow even at unequal time (within the support of
the equal $\tau$ wave functions), as in the proposed experiment of
Palacios {\it et al} [1]. This argument can be followed
for arbitrary $N$,
and therefore the Fock space of quantum field theory, as we show 
below, carries
the properties usually associated with fermion (or boson) fields, with
the entire Fock space foliated over the orbit of the inducing vector
$n^\mu$.
\par Let us now construct a two body Hilbert space in the
framework of the relativistic quantum theory. The states of this two
body space are given by linear combinations over the product wave
functions, where the wave functions (for the spin $(1/2)$ case; the formulation is the same for bosons) are
 of the type described in $(2.27)$,  {\it i.e.} (for equal $n$ and $\tau$),

$$ \psi_{ij}(x_1,x_2) = \psi_i(x_1) \times \psi_j(x_2), \eqno(3.1)$$
where $ \psi_i(x_1)$ and $\psi_j(x_2)$ are elements of the
one-particle Hilbert space ${\cal H}$. Let us introduce the notation,
often used in differential geometry, that
$$ \psi_{ij}(x_1,x_2) = \psi_i \otimes \psi_j (x_1, x_2), \eqno(3.2)$$
identifying the arguments according to a standard ordering. Then,
without specifying the spacetime coordinates, we can write
 $$ \psi_{ij} = \psi_i \otimes \psi_j , \eqno(3.3)$$
formally, an element of the tensor product space ${\cal H}_1 \otimes
{\cal H}_2$. The scalar product is carried out by pairing the elements
in the two factors according to their order, since it corresponds to
 integrals over $x_1, x_2$, {\it i.e.},
$$(  \psi_{ij},  \psi_{k,\ell}) = (\psi_i, \psi_k)(\psi_j, \psi_\ell).
\eqno(3.4) $$
\par For two identical particle states satisfying
Bose-Einstein of Fermi-Dirac statistics, we must write, according to
our argument given above,
$$ \psi_{ijn}= {1 \over \sqrt 2}[\psi_{in} \otimes \psi_{jn} \pm
\psi_{jn} \otimes \psi_{in}], \eqno(3.5)$$
where $n \equiv n^u$ is the timelike four vector labelling the orbit of
the induced representation. This expression has the required symmetry
or antisymmetry only if both functions are on the same points of their
respective orbits in the induced representation. Furthermore, they
transform under the {\it same}  $SU(2)$ representation of the rotation
subgroup of the Lorentz group, and thus for spin $1/2$ particles,
under a $\pi$ spatial rotation (defined by the space orthogonal to the timelike
vector $n^\mu$)
they both develop a phase factor $e^{i{\pi \over 2}}$.  The product
results in an over all negative sign.  As in the usual quantum theory,
this rotation corresponds to an interchange of the two particles, but
here with respect to a ``spatial'' rotation around the timelike vector $n^\mu$. The
spacetime coordinates in the functions are rotated in this (foliated)
subspace of spacetime, and correspond to an actual exchange of the
positions of the particles in space time, as in the formulation of
the standard spin-statistics theorem. It therefore
follows that the interchange of the particles occurs in the foliated
space defined by $n^\mu$, and, furthermore: 
\bigskip 
\noindent {\it The antisymmetry of identical spin $1/2$ (fermionic)
particles, at equal $\tau$, remains at unequal times (within the support of the wave
functions). This is true for the symmetry of identical spin zero
(bosonic) particles as well}.
\bigskip
\par  The construction we have given enables us to define
the spin of a many body system, even if the particles are relativistic
and moving arbitrarily with respect to each other.
\bigskip 
\noindent{\it The spin of an $N$-body system is well-defined, independent of
the state of motion of the particles of the system, by the usual laws
of combining representations of $SU(2)$, i.e, with the usual
Clebsch-Gordan coefficients, if the states of all the particles in the
system are in induced representations at the same point of the orbit
$n^\mu$ and equal} $\tau$.[14]
\par Furthermore, as we have pointed out, the generators of the rotation groups in the fiber $n$ of the foliation, act in the spacelike subspace orthogonal to $n^\mu$. Therefore, orbital angular momenta can as well be combined using standard Clebsch-Gordan addition for any number of particles, independently of the fact that they are in relative motion
\bigskip
\noindent{\it 4. The Palacios et al experiment}
\bigskip
\par The Palacios {\it et al} prediction for the measurment of existence of entanglement of spin $1/2$ elctrons emitted by double ionization of helium rests on the interference that can be observed for the space-time configuration part of the wave functions, which are symmetric, since the spin part is antisymmetric in the spin zero state.  As we have pointed out, the antisymmetry of the spin state at {\it unequal} times (within the support of the wave function) is valid in the SHP theory, and the corresponding spacetime parts of the wave function will be, in the same way,  symmetric.  This experiment would then show interference between parts of the wave function carrying different values of the $t$ variable in the same way as in the Lindner {\it et al} experiment. The orders of magnitude of time intervals in the Palacios {\it et al} configuration are, in fact, due to the characteristic properties of helium, very close to  those of the Lindner {\it et al} experiment. The time intervals involved are therefore also of the order of femtoseconds. Our discussion, in the framework of the SHP theory, considers the time $t$ as an observable, with a spread (rigorously obeying the uncertainty relation $\Delta t \Delta E \geq{1 \over 2} \hbar$) in the wavepackets (on the Hilbert space over the measure $d^4 x$), {\it for both particles at equal} $\tau$.   
\par The two entangled electrons are considered to be emitted, with the same polarization, with energies $E_1 = 35 \  eV$ and $E_2 = 69 \   eV$ (about $10.4 \ eV$ and $14.6 \ eV$ after atomic physics corrections), separated by a time intervals of the order of $.75 \  fs$ (femtoseconds), with emission pulse widths of the order of $0.5 \  fs$ (non-overlapping), here necessarily within the time width of the two-body wave packet. Since this time interval is of the order of the time intervals in the Lindner {\it et al} experiment in the emission of a single electron, the structure of the two-body wave packet should have similar spread in time, the characteristic uncertainty in energy determined by the atomic decay mechanism. As we have remarked in our study [6] of the  Lindner {\it et al} experiment, Floquet theory [22] (for which the time $t$ becomes an observable in a nonrelativistic framework) then would not explain the interference.
\par As formulated by Palacios {\it et al}, the antisymmetric spin zero state is antisymmetric in the spin factors and therefore symmetric in the spacetime factors in the two-body state. We write the spacetime factor for the wave function with both functions in the same foliation sheet $n^\mu$ (we suppress the normalization factor $1/\sqrt{2}$)
$$ \eqalign{\Psi &= \varphi_1(x_1) \varphi_2(x_2) + \varphi_1(x_2) \varphi_1(x_1)\cr
&\cong \varphi_1 (k_1) \varphi_2 (k_2)  \bigl[ e^{i({\bf k_1}\cdot {\bf x_1}+{\bf k_2}\cdot {\bf x_2}- E_1t_1 -E_2 t_2)}\cr
 &+ e^{i({\bf k}_1\cdot {\bf x}_2+{\bf k}_2\cdot {\bf x}_1- E_1t_2 -E_2 t_1)} \bigr], \cr} \eqno(4.1)$$
where we have interchanged the spacetime locations of the two identical electrons in the symmetrization (equivalent to interchange of the states). The two states, $\varphi_1$ and $\varphi_2$ differ in that in the first an electron is emitted from $He$, the second, the second electron is emitted from $He^+$.
\par We now define
$$ T= {t_1 +t_2 \over 2} \eqno(4.2)$$
and
$$ \Delta t = t_2 - t_1, \eqno(4.3)$$
so that Eq.$(3.1)$ becomes
$$ \eqalign{\Psi &= \varphi_1(x_1) \varphi_2(x_2) + \varphi_1(x_2) \varphi_1(x_1)\cr
&\cong \varphi_1 (k_1) \varphi_2 (k_2) e^{-i(E_1 +E_2)T} \bigl[ e^{i({\bf k}_1\cdot {\bf x}_1+{\bf k}_2\cdot {\bf x}_2-{i \over 2}(E_2 - E_1)\Delta t) } \cr
&+ e^{i({\bf k}_1\cdot {\bf x}_2+{\bf k}_2\cdot {\bf x}_1- {i \over 2}(E_1 - E_2)\Delta t) }  \bigr], \cr} \eqno(4.4)$$
in agreement with the structure found by Palacios {\it et al}. (we assume equal pulse widths as in their work). 
\par Carrying out the integrals of the wave packets $\varphi_1 (k_1), \varphi_2 (k_2)$ (here, $E_1 , E_2$ are independent of $k_1,k_2$ ), there will be an additional phase (as in the Palacios {\it et el} calculation, but the $\Delta t$-dependent phase is proportional to $E_2 - E_1$. We remark that these energies, corresponding to the spectra 
of the relativistic atomic bound state problem [23] contain to first order the terms $M_i c^2$ plus the Schr\"odinger, with additional relativistic corrections (here negligible). The $M_i c^2$ terms cancel for two electrons, and the remaining bound state level values would be in agreement with the Palacios {\it et el} calculation.
\bigskip
\noindent{\it 5. Conclusions}
\bigskip
\par We have discussed spin and orbital angular momentum representations in a consistent relativistic quantum theory, generalizing Wigner's construction for the representation of relativistic spin from a foliation over momentum to a foliation over an arbitrary timelike vector $n^\mu$ [12]normalized to  unity. This formulation admits the construction of representations of relativistc spin and angular momentum in a quantum mechanical Hilbert space for which the generators of both spin and anglular momentum act in a spacelike surface orthogonal to the timelike vector $n^\mu$. The standard Clebsch-Gordan methods are applicable to the reduction of direct product representations of two body (or more) states [14], in the fiber labelled by $n^\mu$, and in particular, to relativistic entanglement. The construction of such a state, involving linear combinations of direct products of wave functions {\it at equal} $\tau$, admit correlations are unequal times since the wave functions have support on both space and time (as we have remarked, in practice it is not possible to prepare a two-body state at precisely equal times).
\par Since the pulse spacings assumed by  Palacios {\it et el} were about $0.75 fs$, interference would be supported between the two two-body states in superposition with wave function widths of this order of magnitude. 
The uncertainty relation then implies that $\Delta E  \geq 10^{-3} eV$. Natural line widths in atomic physics appear to be of order $10^{-6} eV$, so that the uncertainty in time in the Stueckelberg wave packet could be much larger than what is needed to account for the observation of interference in time in the entangled state.   
\bigskip
\bigskip

\centerline{\bf References}
\bigskip
\frenchspacing
{\obeylines\smallskip
\item{1.}  A. Palacios, T.N. Rescigno and C.W. McCurdy,
Phys. Rev. Lett. {\bf 103} 253001 (2009).
\item{2.} F. Lindner, M.G. Sch\"atzel, H. Walther, A.Baltuska,
E. Goulielmakis, F. Krausz, D.B. Milo\v sevi\'c, D. Bauer, W. Becker
and G.G. Paulus, Phys. Rev. Lett.{\bf 95}, 040401 (2005).
\item{3.}  G. Ludwig, {\it Foundations of Quantum Mechanics
I},Springer-Verlag, New York (1982); {\it Foundations of Quantum
Mechanics II},Springer-Verlag, New York (1983); P.A.M. Dirac,{\it Quantum Mechanics}, First edition, Oxford Univ. Press, London (1930), third edition (1947).;  P.A.M. Dirac, Proc. Roy. Soc. (London) {\bf A 136}, 453 (1932).
\item{4.}  G. Baym, {\it Lectures on Quantum Mechanics},
W.A. Benjamin, N.Y. (1969); A.L. Fetter and J.D. Walecka, {\it Quantum
Theory of Many Particle Systems}, McGraw Hill, New York (1971).
\item{5.}  E.C.G. Stueckelberg, Helv. Phys. Acta {\bf
14}, 372, 585; {\bf 15}, 23 (1942);  L.P. Horwitz and C. Piron, Helv. Phys. Acta {\bf 66}, 316
(1973). See also R.E. Collins and J.R. Fanchi, Nuovo Cim. {\bf 48A},
314 (1978);  J.R. Fanchi, {\it Parametrized Relativistic Quantum
Theory}, Kluwer, Dordrecht (1993). We remark that the this theory was developed, as emphasized in these references, in order to provide a framework for solving the problems raised by, for example, L.H. Thomas, Phys. Rev. {\bf 85}, 868 (1952), H. Van Dam and E.P.Wigner, Phys. Rev. {\bf 85}, 868 (1965),  and D.G. Currie, T.F. Jordan and E.C.G. Sudarshan, Rev. Mod. Phys. {\bf 35}, 350 (1963). We wish to thank Profs. Sudarshan and Jordan, in particular, for discussions.   
\item{6.} L.P.  Horwitz, Phys. Lett.  {\bf A 355}, 1  (2006).
\item{7.}  L.P. Horwitz and Y. Rabin, Lettere al Nuovo
Cimento {\bf 17} 501 (1976).
\item{8.}  E.P. Wigner, Phys. Rev. {\bf 98}, 145 (1955).
\item{9.}  G.W.  Mackey, {\it Induced Representations of Groups and
Quantum Mechanics}, Benjamin, New York (1968); V. Bargmann, Ann. of
Math. {\bf 48}, 568 (1947).}
\item{10.} A. Clebsch, {\it Theorie der bin\"aren
algebraischen Formen}, Teubner, Leuipzig (1872); P. Gordan,
{\it \"Uber das Formensystem bin\"arer Formen}, Teubner, Leipzig
(1875). See also, A.R. Edmunds, {\it Angular Momomentum in Quantum
Mechanics}, Princeton Univ. Press, Princeton (1957); L.C. Biedenharn
and J.D. Louck, {\it Angular Momentum in Quantum Physics, Theory and 
Application}, Encyclopedia of Mathematics and its Applications,
Ed. Gian-Carlo Rota, vol. 8,  Addison Wesley, Reading, Mass.(1981).
\item{11.}  A. Einstein, B. Podolsky and N. Rosen,
Phys. Rev. {\bf 48}696 (1935).
\item{12.}  L.P. Horwitz, C. Piron and F. Reuse,
Helv. Phys. Acta {\bf 48}, 546 (1975). See also, Lawrence P. Horwitz, {\it Relativistic Quantum Mechanics}, Fundamental Theories of Physics {\bf 180}, Springer, Dordrecht (2015).
\item{13.} R. Arshansky and L.P. Horwitz,J. Phys. A:
Math. Gen. {\bf 15}L659 (1982).
\item{14.} Lawrence P. Horwitz and Meir Zeilig-Hess, Jour. Math. Phys.{\bf 56} 002301 (2015); see also, L.P. Horwitz, Jour. Phys. A:Math and Gen {\bf 46}, 035305 (2013).
\item{15.}  H. Boerner, {\it Representations of Groups}, p.312,
North Holland, Amsterdam (1963).
\item{16.} J.D. Bjorken and S.D. Drell, {\it Relativistic
Quantum Mechanics}, McGraw Hill, New York (1964).
\item{17.}  S. Weinberg, {\it Quantum Field Theory I}, p. 49,
Cambridge Univ. Press, Cambridge (1995).
\item{18.}  J. Schwinger, Physical Review {\bf 82}, 664
 (1951).
\item{19.}  A. Bennett, Jour. Phys. A: Math. Theor. {\bf
45} 285302 (2012). 
\item{20.}  Lawrence Horwitz, Jour. Pys. A: Math and Theor. {\bf
46} 035305 (2013).
\item{21.}  J. Schwinger, Phys. Rev. {\bf 73}, 416 (1948); Phys. Rev. {\bf 74}
1439 (1948); S. Tomonaga, Phys. Rev. {\bf 74} 224 (1948);
\item{22.} M.G. Floquet, Ann. Ecole Normal Suppl. {\bf 12}, 47 (1883). See also, J.S. Howland, Math. Ann.{\bf 207} 315 (1974); Indiana Math. J. {\bf 28} 471 (1979).
\item{23.} R.I. Arshansky and L.P. Horwitz, Jour. Math. Phys. {\bf 30},
66, 380 (1989); R.I. Arshansky and L.P.Horwitz,
Jour. Math. Phys. {\bf 30}, 213 (1989).

\end